\documentclass[]{aa} 

\usepackage[varg]{txfonts}
\usepackage{graphicx}

\begin{document}

\title{Pulsations powered by hydrogen shell burning in white dwarfs} 

\author{M. E. Camisassa\inst{1,2} \and
        A. H. C\'orsico\inst{1,2} \and    
        L. G. Althaus\inst{1,2} \and
        H. Shibahashi\inst{3}}
\institute{Grupo de Evoluci\'on Estelar y Pulsaciones. 
           Facultad de Ciencias Astron\'omicas y Geof\'{\i}sicas, 
           Universidad Nacional de La Plata, 
           Paseo del Bosque s/n, 1900 
           La Plata, 
           Argentina 
           \email{camisassa@fcaglp.unlp.edu.ar}
           \and
           Instituto de Astrof\'isica de La Plata, Centro Cient\'ifico Tecnol\'ogico La Plata, 
           Consejo Nacional de Investigaciones Cient\'ificas y T\'ecnicas,
           Paseo del Bosque s/n, 1900 
           La Plata, 
           Argentina
           \and
           Department of Astronomy, The University of Tokyo, 
           Bunkyo-ku, Tokyo 113-0033, Japan}
\date{Received ; accepted }

\abstract{
In the absence of a third dredge-up episode 
during the asymptotic giant branch phase, white dwarf models evolved from low-metallicity progenitors have a 
thick hydrogen envelope, which makes  hydrogen shell burning be 
the most important energy source.}
{
We investigate the pulsational stability of white dwarf models with thick envelopes
to see whether nonradial $g$-mode pulsations are triggered by hydrogen  
burning, with the aim of placing constraints on hydrogen  
shell burning in cool white dwarfs and on  
a third dredge-up during the asymptotic giant branch 
evolution of their progenitor stars.}
{
We construct white-dwarf  sequences  from low-metallicity 
progenitors by means of full evolutionary 
calculations that take into account the entire history    of  progenitor
stars,     including    the thermally pulsing  and  the
post-asymptotic  giant  branch phases,  and analyze their 
pulsation stability by solving the linear, nonadiabatic, nonradial 
pulsation equations for the models in the range of effective temperatures 
$T_{\rm eff} \sim 15\,000\,-\, 8\,000$ K. 
}
{
We demonstrate that, for white dwarf models with masses  
$M_{\star} \lesssim 0.71\,\rm M_{\sun}$ and effective temperatures 
$8\,500 \lesssim T_{\rm eff} \lesssim 11\,600$ K that evolved from low-metallicity 
progenitors ($Z= 0.0001$, $0.0005$, and $0.001$), 
the dipole ($\ell= 1$)  and quadrupole ($\ell=2$) $g_1$ modes are excited mostly due to
the hydrogen-burning shell through the $\varepsilon$-mechanism,  
in addition to other $g$ modes driven by 
either the  $\kappa-\gamma$  or the convective driving mechanism.
However, the $\varepsilon$ mechanism is insufficient to drive 
these modes in white dwarfs evolved from solar-metallicity progenitors.} 
{
We suggest that efforts should be made to observe the dipole 
$g_1$ mode  in white dwarfs associated with low-metallicity environments, 
such as globular clusters and/or the galactic halo, 
to place constraints on  hydrogen shell burning in cool white dwarfs and the  
third dredge-up episode during the preceding asymptotic giant branch phase.
}
\keywords{stars: evolution --- stars: interiors --- stars: oscillations --- white dwarfs}
\titlerunning{Pulsations powered by hydrogen shell burning in white dwarfs}
\authorrunning{Camisassa et al.}  

\maketitle


\section{Introduction}
\label{introduction}

White dwarf stars are the most common end product of stellar
evolution, with 97\% of all stars becoming white dwarfs.  As a result,
observations of white dwarfs provide us  rich information about the
stellar evolution of their progenitors and the star formation history
in our galaxy.  Furthermore, some white  dwarfs are pulsating stars,
hence asteroseismology of these stars  makes further information
available about the internal structure and the evolutionary status of
white dwarfs.  Several types of pulsating white dwarfs have been
observed:  GW Vir (pulsating PG1159) stars, hot DA variable (HDAV)
stars,  V777 Her (DBV) stars, hot DQ variable (HDQV) stars, ZZ Ceti
(DAV) stars,  and Extremely Low Mass variable (ELMV) stars
\citep{2008ARA&A..46..157W,2008PASP..120.1043F,
  2010A&ARv..18..471A,2013ASPC..479..211F}.  They exhibit luminosity
variations with periods ranging from $\sim\!100$\,s to
$\sim\!6000$\,s, which are attributed to nonradial $g$ modes   excited
by either the  $\kappa-\gamma$ mechanism that works in the ionization zones of
the abundant chemical elements  or convection in cool white dwarfs
\citep{1983MNRAS.204..537B,1999ApJ...511..904G}. 

Besides those mechanisms, the $\varepsilon$ mechanism has been
predicted to lead to destabilization of some short-period $g$ modes in
white-dwarf and pre-white-dwarf stars. This mechanism is  induced by
thermonuclear reactions, which are highly dependent on temperature.
During a compression phase, the temperature and thus the nuclear
energy  production rates are higher than at equilibrium, hence,  in
the layers where nuclear reactions take place, more thermal energy is
produced. The opposite occurs during the following expansion phase,
and as a consequence, the contribution of the perturbation of nuclear
energy generation to the work integral is always a destabilization
effect \citep{1989nos..book.....U}. The first study of the
$\varepsilon$ mechanism  in   pre-white-dwarf stars was the seminal
work by  \citet{1986ApJ...306L..41K},  who found that some $g$ modes
with periods in the range $70-200$\,s are indeed excited  through this
mechanism by  helium shell burning in hydrogen-deficient
pre-white-dwarf stars.   Later on, \citet{2009ApJ...701.1008C}
conducted a more thorough stability analysis, and showed the existence
of an instability strip for  GW Vir stars in the H-R diagram, for
which the $\varepsilon$ mechanism by helium shell burning is
responsible.  As for hydrogen burning,  \citet{2014PASJ...66...76M}
recently predicted that some low-order  $g$ modes are excited by the
$\varepsilon$ mechanism due to vigorous hydrogen  shell burning in
very hot and luminous   pre-white dwarfs with hydrogen-rich envelopes.
Also, \citet{2014ApJ...793L..17C} have shown that some low-order $g$
modes  could be driven in ELMVs by the $\varepsilon$ mechanism due to
hydrogen shell burning, although most of the observed pulsation modes,
which are characterized by high- and intermediate- radial orders, are
likely to be excited by the  $\kappa-\gamma$ mechanism. 

Recent detailed studies by \citet{2013ApJ...775L..22M} and
\citet{2015A&A...576A...9A} have shown that   white dwarf stars
evolved from low-metallicity  progenitors  ($0.00003\lesssim Z
\lesssim 0.001$), in the absence of  carbon  enrichment due to third
dredge-up episodes during the  asymptotic giant branch (AGB) phase,
have a thick enough hydrogen envelope, which makes stable hydrogen
shell burning by pp-chain the most important energy source even at low
luminosities  ($\log L/{\rm L}_{\odot}\lesssim-3$). 
Nuclear burning becomes weaker with an increase of
metallicity.   The importance of  hydrogen shell burning is also
strongly dependent on the mass of the white dwarf, and   no
significant burning is expected to occur in white dwarfs with masses
greater than $\sim 0.6\, {\rm M}_{\sun}$.     This extra source of
energy delays the white dwarf cooling.  Even at low luminosities,
$\log L/{\rm L}_{\odot}\sim-4$, residual hydrogen burning leads to an
increase in the cooling times of the low- and  intermediate-mass white
dwarfs  ($M\lesssim 0.6\, {\rm M}_{\sun}$) by $\sim 20-40\%$.  
These results are not affected by the mass
loss rate  in the preceding stages. On the other hand, the occurrence
of overshooting during the AGB phase strongly favors third dredge-up
episodes.  This leads to the formation of white dwarfs with thinner hydrogen envelopes,
with the consequence that no appreciable residual hydrogen burning during 
the white dwarf regime is expected.
The delay of the cooling times of white dwarfs evolved
from low-metallicity progenitors impacts on our
understanding of stellar populations. In particular,
\citet{2015A&A...581A..90T} employed  the observed white dwarf
luminosity function of the low-metallicity globular cluster NGC\,6397
to constrain the hydrogen nuclear burning in white dwarfs from
low-metallicity progenitors. They conclude that the  low-mass white
dwarf population in NGC\,6397  should be characterized by important
residual burning  and hence, the progenitors of these stars should
not have experienced third dredge-up in the AGB phase.  Although
\citet{2015A&A...581A.108C} tried the same with the isolated white
dwarfs in the galactic halo, they could not  reach a definitive
conclusion,  because the number of  detected   white dwarfs in the
galactic halo remains too small. 

According to \citet{2013ApJ...775L..22M} and
\citet{2015A&A...576A...9A}, the  contribution of nuclear energy
generation to the luminosity of the white dwarfs evolved from
low-metallicity progenitors is the largest, being about $80\%$ of the
luminosity of the star, when $\log T_{\rm eff}\sim 4$.  It is in this
effective temperature range that  many DAVs are found to pulsate in
$g$ modes driven by  either the  $\kappa-\gamma$ mechanism or the convective
driving mechanism.   Motivated by these results, in this paper  we
assess the possibility that DA white dwarfs  belonging to
low-metallicity stellar populations (like globular  clusters and the
galactic halo) can  develop short-period variabilities, analogous to
the short-period $g$ modes  exhibited by ZZ Ceti stars. As described
above, the low-mass and low-metallicity progenitors  of this kind of
DA white dwarfs should avoid the third dredge-up  episodes in the AGB
phase, such that residual hydrogen burning continues at a significant
level, and  overstability of  $g$ modes is induced  by the
$\varepsilon$ mechanism.  Along this study, we aim to use white dwarf
pulsations to verify the presence of   hydrogen shell burning in some
white dwarfs, and eventually,  to constrain the efficiency of third
dredge-up episodes in the  AGB phase.  

The paper is organized as follows: In Sect. \ref{modeling}  we
briefly describe our   stellar evolution and stellar pulsation tools
and the main ingredients of  the evolutionary sequences. In
Sect. \ref{results} we present our  pulsation results in
detail. Finally, in Sect. \ref{conclusions}  we summarize our main
findings.

\section{Evolutionary models and numerical tools}
\label{modeling}

We have constructed equilibrium models of low-mass 
white dwarfs evolved from low-metallicity progenitors, 
using the {\tt LPCODE} stellar evolutionary code
\citep{2003A&A...404..593A, 2005A&A...435..631A, 2012A&A...537A..33A, 2015A&A...576A...9A, 2016ApJ...823..158C},
which were initially evolved from the zero
age main sequence (ZAMS), all the way to the white-dwarf stage.
Additionally, we computed an evolutionary sequence of solar metallicity
($Z= 0.018$) in order to see if the $\varepsilon$ mechanism  
could also excite oscillations in white dwarfs coming from
solar-metallicity progenitors. The physical inputs of our models 
were described in detail by \citet{2015A&A...576A...9A}.
Below, we list the main ingredients 
which are relevant for this work: 

\begin{itemize}
\item Extra mixing  due to  diffusive convective  overshooting has
been considered during  the core hydrogen and helium 
burning phases.  Following \citet{2015A&A...576A...9A}, extra mixing was not considered 
during the thermally-pulsing  AGB. The
Overshooting parameter was set to $f=0.015$ at each convective boundary.
\item  Mass loss  during the red giant branch phase was taken  from \citet{2005ApJ...630L..73S}.
For the AGB phase, we use again that of \citet{2005ApJ...630L..73S} for
pulsation periods shorter than 50  days. For longer periods, mass loss
is taken as the maximum of the rates of \citet{2005ApJ...630L..73S} and
\citet{2009A&A...506.1277G} for  oxygen rich  stars, or the  maximum of
the rates of \citet{2005ApJ...630L..73S} and \citet{1998MNRAS.293...18G}
for  carbon-rich stars.   In all  of our  calculations, mass  loss was
suppressed after the  post-AGB remnants reach $\log T_{\rm  eff} = 4$.

\item Radiative   and    conductive   opacities   are   taken    from   OPAL
\citep{1996ApJ...464..943I}   and   from   \citet{2007ApJ...661.1094C},
respectively. For the low-temperature regime we used
molecular  opacities with varying  carbon to oxygen  ratios 
\citep{2005ApJ...623..585F, 2009A&A...508.1343W}.
These opacities are necessary for  a  realistic treatment
of progenitor  evolution during  the thermally-pulsing  AGB phase. 
\item For  the  white-dwarf  stage  we  employed an updated 
version of  the  equation  of  state  of \citet{1979A&A....72..134M} 
for  the low-density regime, whereas  for the high-density regime we 
considered the equation of state of \citet{1994ApJ...434..641S}.
\item Changes in the chemical abundances have been computed 
consistently with the predictions of 
nuclear burning and convective mixing all the way from the ZAMS to the white
dwarf phase. In addition, abundance changes resulting from atomic element 
diffusion were considered during the white dwarf regime.
\item For  the white-dwarf stage and for effective temperatures 
lower than 10,000\,K, outer boundary conditions for the evolving  
models are derived from non-gray model  atmospheres
\citep{2012A&A...546A.119R}.
\end{itemize}

Table\,\ref{table} lists the initial masses of the sequences, 
the resulting white dwarf masses, the hydrogen mass fraction,
$\log{(M_{\rm H}/{\rm M}_{\sun})}$, and the surface
 carbon to oxygen ratio, C/O, at the beginning of the white dwarf 
phase. 
 Our prediction for the initial-to-final mass relation
is similar to that calculated by \citet{2015MNRAS.450.3708R}.
The C/O ratio indicates that, with 
the exception of the $M_{\star}= 0.738\,{\rm M}_{\sun}$ sequence with 
$Z= 0.0001$, none of these model stars experienced carbon enrichment 
of the envelope during the AGB phase. The total amount of hydrogen 
at the beginning of the white dwarf stage ($M_{\rm H}$) is a key quantity
for assessing the importance of nuclear burning, since the more massive
the hydrogen envelope is,  the more intense the burning 
is.  In this work, we have not considered that some white dwarfs could have thin
hydrogen envelopes \citep{2008MNRAS.385..430C,2009MNRAS.396.1709C,2012MNRAS.420.1462R,2013ApJ...779...58R}

\begin{figure*}
\centering
\includegraphics[clip,width=17cm]{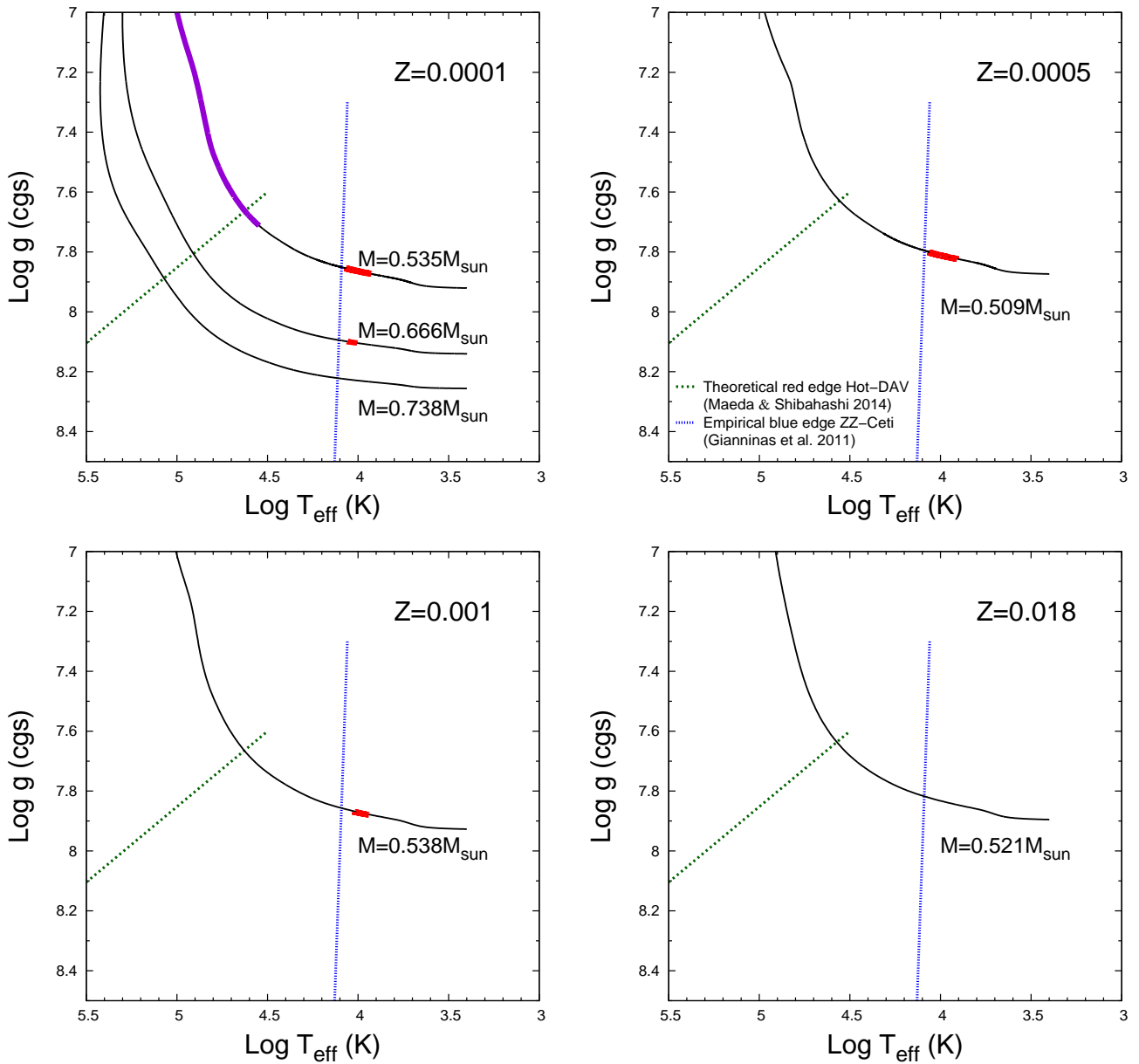}
\caption{Evolutionary tracks of white dwarf
models with $M_{\star}\sim 0.51$\,-\,$0.54\,\rm M_{\sun}$ on the 
$\log T_{\rm eff}$\,-\,$\log g$ plane. Each panel corresponds to 
a different metallicity value ($Z$) of the white dwarf progenitors. 
For the case of $Z= 0.0001$ we include 
two additional sequences with masses $M_{\star}= 0.666{\,}\rm M_{\sun}$  
and $M_{\star}= 0.738{\,}\rm M_{\sun}$. Red thick segments on the tracks at 
$\log T_{\rm eff} \sim 4$ correspond to DA white dwarf models having
the dipole  $g_1$ mode destabilized mostly by the $\varepsilon$ mechanism
acting at the hydrogen burning shell. The violet thick segment
in the case of the $M_{\star}= 0.535\,\rm M_{\sun}$ and $Z= 0.0001$
sequence corresponds to models with unstable low-order
$g$-modes ($g_1$ to $g_4$) destabilized solely by the $\varepsilon$
mechanism. The dashed green line displays the theoretical red edge of the 
instability domain of pre-white-dwarfs with hydrogen-rich
envelopes evolved from $M_{\rm ZAMS}\ge 1.5\,{\rm M}_\odot$ with
solar metallicity found by \citet{2014PASJ...66...76M}. The almost
vertical dotted blue line corresponds to the empirical blue edge
of the instability domain of ZZ Ceti stars reported by
\citet{2011ApJ...743..138G}.}
\label{Fig:logtlogg}
\end{figure*}

The pulsation computations were performed with the linear, nonradial,
nonadiabatic version of the {\tt LP-PUL} pulsation code
described by \citet{2006A&A...458..259C} \citep[see also][]
{2009ApJ...701.1008C}.  Although we have considered both dipole ($\ell= 1$) 
and quadrupole ($\ell= 2$) $g$ modes,  
the results  are qualitatively similar and therefore
we will focus on results for $\ell= 1$. The pulsation
code solves the sixth-order complex system of linearized equations and
boundary  conditions as given by \cite{1989nos..book.....U}. The code
considers  the mode excitation by both the  $\kappa-\gamma$ mechanism
and the $\varepsilon$ mechanism. Our computations ignore  the
perturbation of the convective flux, that is,  we are adopting 
the ``frozen-in convection'' approximation. 
While this approximation is known to give unrealistic locations 
of the g-mode red edge of the ZZ Ceti instability strip, it leads to satisfactory 
predictions for the location of the blue edge \citep{2012A&A...539A..87V,2013EPJWC..4305005S}. The 
same is true for the blue edge of the instability domain of the ELMV stars
\citep{2016A&A...585A...1C}. We have also performed
additional calculations disregarding the effects  of nuclear energy
release on the nonadiabatic pulsations, for which  we fix 
$\varepsilon= \varepsilon_{\rho}= \varepsilon_{T}= 0$,
where $\varepsilon$ is the nuclear energy production rate,
and  $\varepsilon_{\rho}$ and $\varepsilon_{T}$ are their
logarithmic derivatives, defined as
$\varepsilon_{\rho}=\left(\partial \ln \varepsilon/\partial \ln \rho \right)_{T}$ 
and
$\varepsilon_{T}= \left(\partial \ln \varepsilon/\partial \ln T\right)_{\rho}$, 
respectively.  In this way,  we prevent the
$\varepsilon$ mechanism from operating, although nuclear burning was
still taken into account in the  evolutionary computations.

While the temperature dependence of energy
generation all through the pp-chain in equilibrium
is governed by the slowest reaction between two protons and
$\varepsilon_T\simeq 4$, we adopted the effective temperature dependence
of nuclear reaction through the perturbation, which
is mainly governed by $^3$He-$^3$He reaction and $\varepsilon_T\simeq 11$
\citep{1989nos..book.....U}.

\begin{table}
\caption{Basic  model  properties   for  sequences  with  Z=0.0001,
  0.0005, 0.0010 and 0.0180.}  \centering
\begin{tabular}{lcccc}
\hline
\hline
$Z$ & $M_{\rm ZAMS}(\rm M_\odot)$ & $M_{\rm WD} (\rm M_\odot)$ & $\log(M_{\rm H}/\rm M_\odot)$ & C/O \\
\hline
$0.0001$ & $0.85$ & $0.535$ & $-3.304$ & $0.301$ \\
$0.0005$ & $0.80$ & $0.509$ & $-3.231$ & $0.241$ \\
$0.0010$ & $0.85$ & $0.538$ & $-3.434$ & $0.307$ \\
$0.0180$ & $1.00$ & $0.521$ & $-3.543$ & $0.005$ \\
$0.0001$ & $1.50$ & $0.666$ & $-3.851$ & $0.264$ \\
$0.0001$ & $2.00$ & $0.738$ & $-4.347$ &  $11.477$ \\
\hline
\end{tabular}
\tablefoot{$M_{\rm  ZAMS}$: initial  mass, $M_{\rm  WD}$: white  dwarf
  mass, $\log  (M_{\rm H}/\rm M_\odot)$: logarithm of  the mass of the hydrogen envelope
  at the maximum effective  temperature  at  the  beginning of  the
  cooling  branch, C/O: surface carbon to oxygen mass ratio at the
  beginning of the cooling branch.}
\label{table}
\end{table}

\begin{figure*}
\centering
\includegraphics[clip,width=16cm]{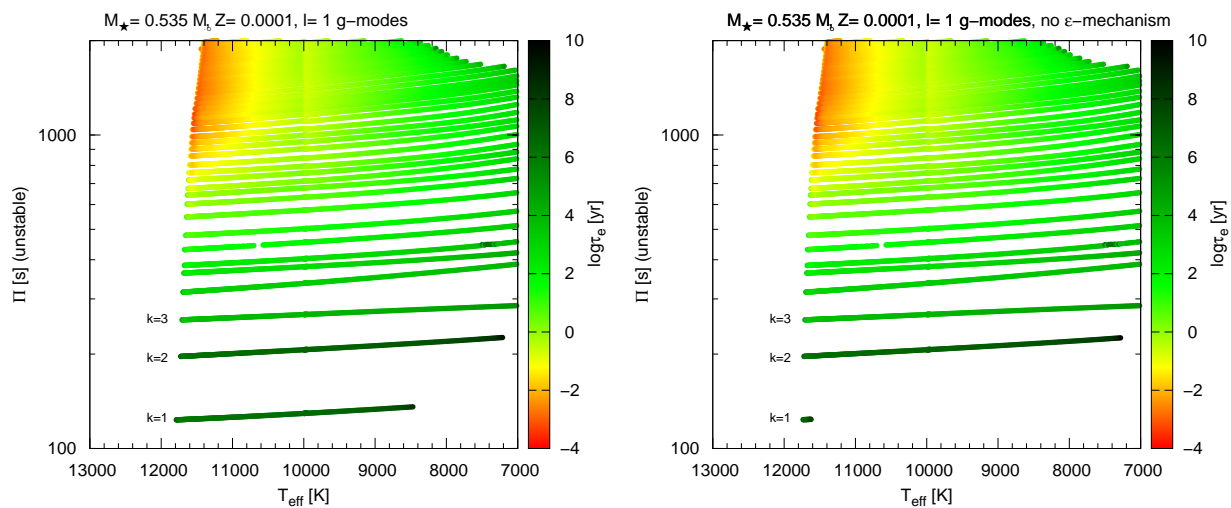}
\caption{Periods ($\Pi$) of unstable dipole $g$ modes  
  in terms of the effective temperature for our white dwarf  model
  sequence with
  $M_{\star}= 0.535{\,}\rm M_{\sun}$ and $Z= 0.0001$.  The left panel corresponds 
  to the sequence in which the $\varepsilon$ mechanism was allowed to operate,
  whereas in the right panel we show the situation in which it was artificially
  suppressed. Color coding indicates  
  the $e$-folding time ($\tau_e$) of each unstable mode (right scale).}
\label{Fig:P0001}
\end{figure*}

\section{Results}
\label{results}

\begin{figure}
\centering
\includegraphics[clip,width=8cm]{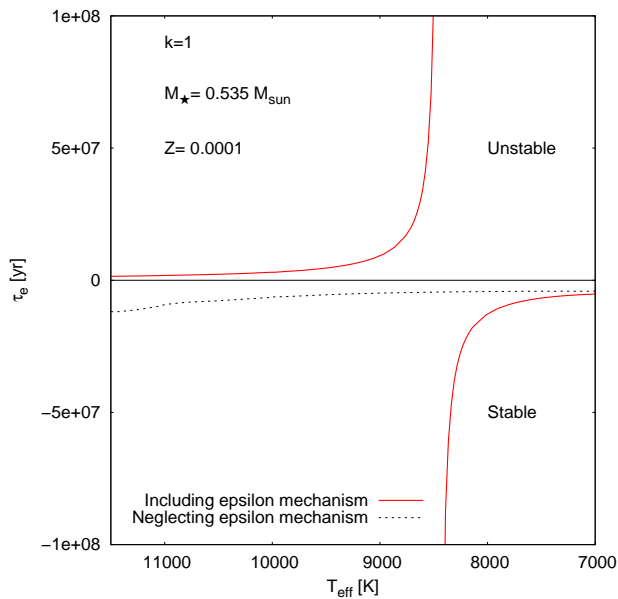}
\caption{$e$-folding time of the $g_1$ mode of
    our $M_{\star}= 0.535{\,}\rm M_{\sun}$ and $Z= 0.0001$
    white dwarf sequence, in terms of the effective temperature.}
\label{Fig:efolding}
\end{figure}

Our stability analysis indicates the existence of  many  unstable
$g$ modes in all the white dwarf model sequences, most of the modes
being  destabilized by the  $\kappa-\gamma$ mechanism acting at the surface hydrogen 
ionization zone. 
 From observational grounds, we know that as the ZZ Ceti stars cool inside the instability strip, modes 
with progressively higher radial orders are excited. This can be understood theoretically on the basis  of
the thermal timescale argument \citep{1980tsp..book.....C,2008ARA&A..46..157W}.
As the star cools, the base of the outer convection zone moves inwards, 
whereby the thermal timescale becomes increasingly higher there. Since the modes that can be excited have 
periods similar to the thermal timescale at the basis of the convection zone, modes with progressively 
longer periods (i.e., with higher radial orders) are excited as the star cools.
When we shut down the $\varepsilon$ mechanism, the
$g_1$ mode becomes stable for the complete range (or part) of
effective temperatures analyzed. This indicates that these modes are
excited to a great extent by the $\varepsilon$ mechanism through the
hydrogen-burning shell.  Higher radial-order modes, on the other hand,
are insensitive to  the effects of nuclear burning, and remain
unstable due only to the  $\kappa-\gamma$ mechanism.

Figure\,\ref{Fig:logtlogg} shows the evolutionary tracks of all our
white  dwarf sequences in the $\log T_{\rm eff}$\,-\,$\log g$ diagram.  Red
thick segments on the evolutionary tracks indicate white dwarf models
with the  dipole  $g_1$ mode excited mostly by the
$\varepsilon$ mechanism.   The figure shows that low-mass white dwarfs
evolved
from progenitor stars  characterized by metallicities
in the range $0.00003 \lesssim Z \lesssim 0.001$ are expected to
undergo $g$-mode pulsations powered by nuclear
burning at  the typical effective temperatures  
of ZZ Ceti stars. Nevertheless,  in white dwarf models derived
from solar metallicity progenitors (right-hand lower panel), which
are indeed  representative of ZZ Ceti stars (which are located in
the solar neighborhood), nuclear burning does not play a role, and
the $\varepsilon$-mechanism is not operative.   For the case of
white dwarf models evolved from progenitors  with $Z= 0.0001$
(left-hand upper panel), we have explored the  efficiency of the
$\varepsilon$-mechanism by changing  stellar masses, and found that
for  models more massive than $\sim 0.71{\,}\rm M_{\sun}$, no $g$ modes
are excited  by nuclear burning. We have included in
Fig.{\,}\ref{Fig:logtlogg} a dotted blue line, which
corresponds to the empirical blue edge of the ZZ Ceti instability
strip  according to \citet{2011ApJ...743..138G}. Clearly,  models
having the pulsationally unstable $g_1$ mode excited by the
$\varepsilon$ mechanism are located in (or near) the ZZ Ceti instability
domain. We remark that the  $g_1$ mode found to be unstable in our
computations is not merely excited by the $\varepsilon$ mechanism,
because the  $\kappa-\gamma$ mechanism contributes somewhat to its destabilization.

Furthermore, we include in Fig.\,\ref{Fig:logtlogg} the location of
the theoretical end of the instability  domain of pre-white-dwarf
models with hydrogen-rich envelope (dashed green line), as  predicted
by \citet{2014PASJ...66...76M}. Note that, at variance with the
$\varepsilon$-destabilized  $g_1$ mode found in this paper in the ZZ
Ceti instability domain, the unstable modes in  the pre-white-dwarf
models considered by \citet{2014PASJ...66...76M} are excited only by
the $\varepsilon$ mechanism. This is because at those high effective
temperatures and luminosities, hydrogen is completely ionized at the
surface layers, hence the  $\kappa-\gamma$ mechanism is inhibited to
operate. We have performed additional stability computations to
investigate whether $g$ modes are excited by the $\varepsilon$ mechanism
at higher luminosities.  For these exploratory computations, we have
considered a template white dwarf  evolutionary sequence with $M_{\star}= 0.535
{\,} \rm M_{\sun}$ and progenitor metallicity $Z= 0.0001$, from
$T_{\rm eff } \sim 100\,000 {\,} \rm K$  downwards (violet thick
line).  We found that dipole modes $g_1$, $g_2$, $g_3$ and $g_4$ are
excited by  the $\varepsilon$ mechanism due to hydrogen burning  down to an
effective temperature of $\sim 36\,000 {\,} \rm K$. These results are
in line with those obtained by \citet{2014PASJ...66...76M}, although
the instability domain obtained here extends to lower effective
temperatures, by virtue that our white dwarf models are characterized
by  more vigorous hydrogen burning. A full exploration of the instability
domain at high luminosities by varying the stellar mass and the
metallicity of the progenitors is, however, beyond the scope of the present
paper, and will be presented in a future publication.

Figure\, \ref{Fig:P0001} shows the unstable periods of dipole 
$g$ modes in terms of the effective temperature for our white dwarf
model sequence with $M_{\star}= 0.535{\,}\rm M_{\sun}$ and $Z= 0.0001$.  The 
left panel shows the results for the standard nonadiabatic pulsation 
computations. A dense spectrum of $g$ modes is destabilized by 
the  $\kappa-\gamma$ mechanism for effective temperatures below
$\sim\!\!11,800$\,-\,$11,400${\,}K. The right  
panel corresponds to the same sequence but for the case in which 
the $\varepsilon$ mechanism is artificially inhibited,
and consequently, only the  $\kappa-\gamma$ mechanism is allowed to act.  
The palette of colors (right-hand scale) indicates  
the $e$-folding time (in years) of each unstable mode,
defined as $\tau_e= 1/|\Im (\sigma)|$, 
where $\Im(\sigma)$ is the imaginary part of the
complex eigenfrequency $\sigma$. The $e$-folding time is a measure of
the time taken for the perturbation to reach an observable amplitude.
For the $M_{\star}= 0.535{\,}\rm M_{\sun}$ and $Z= 0.0001$ sequence, 
the dipole $g_1$ mode is unstable in the standard computations 
(left panel), but the mode becomes stable when we neglect the effect of
nuclear energy production on stability (right panel). Hence, we  
conclude that the mode is unstable due to the $\varepsilon$ mechanism. 
The $e$-folding time of this mode is  shown in terms of the effective
temperature in Figure\, \ref{Fig:efolding}. When the $\varepsilon$
mechanism is allowed to operate (solid red line), the $e$-folding time
is about $5 \times 10^6$ years, whereas the time the star needs to
cross the instability domain is of about $1.2\times10^9$ years.
Therefore, if the nuclear energy production 
is as intense as our models predict, the dipole g$_1$ mode 
has enough time to reach observable amplitudes while the star 
is still evolving in that effective temperature range. 
A similar situation is found for the white dwarf models evolved 
from progenitors with $Z= 0.0005$ and $Z= 0.001$ 
sequences, as shown in the upper right and the lower left panels of Fig.\, \ref{Fig:logtlogg}, respectively.
In these cases, however, the efficiency of  
the $\varepsilon$ mechanism is lower (due to less vigorous nuclear burning), 
and as a result, the ranges in $T_{\rm eff}$ in which the dipole $g_1$ mode
is destabilized by this mechanism are narrower. In the extreme case of 
white dwarfs  evolved from solar-metallicity progenitors, that mode is found to be no longer excited by  
the $\varepsilon$ mechanism.  This result is expected
because, for these white dwarfs, nuclear burning plays a minor role at
advanced stages of their evolution. 

\begin{figure*}
\centering
\includegraphics[clip,width=17cm]{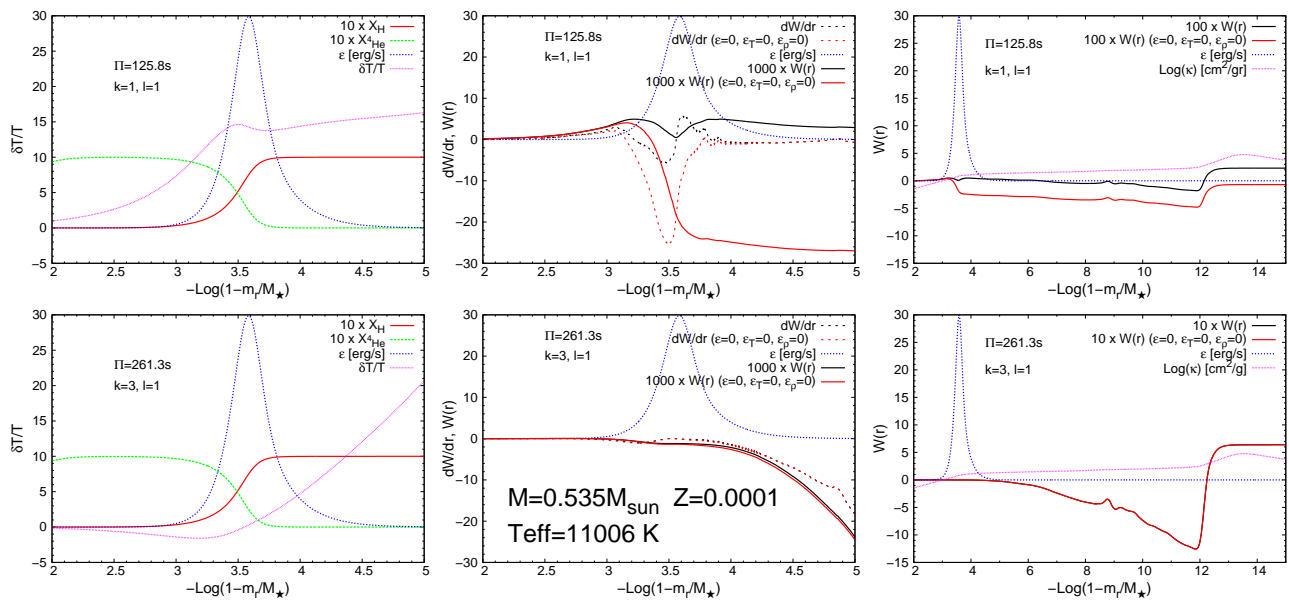}
\caption{Left panels: Lagrangian perturbation of temperature
  ($\delta T/T$), nuclear generation rate ($\varepsilon$),
 and the mass fractions of hydrogen and helium
($X_{\rm H}$ and $X_{\rm He}$, respectively) 
as functions of  $\log(1-M_r/M_{\star})$  
for dipole $g_1$ mode (upper panel) and $g_3$ mode  (lower panel),
respectively,  for the model with  $Z= 0.0001$,  
$M_{\star}= 0.535\,\rm M_{\sun}$ and $T_{\rm eff}= 11,000$ K.  
The eigenfunction is normalized so that the relative displacement
in the radial direction is unity at the stellar surface.
Middle panels: same as left panels but for the differential work 
function (${\rm d}W(r)/{\rm d}r$) for the case in which the
$\varepsilon$ mechanism is allowed to operate (black 
dashed curves) and when it is suppressed (red dashed curves).  Also, 
the scaled running work integrals ($W(r)$) are shown with solid
lines. Right panels: same as central panels, but 
the scaled running work integrals are shown for a wider range of 
the mass coordinate, along with the Rosseland mean opacity 
$\kappa$ (radiative plus conductive).}
\label{Fig:wf}
\end{figure*}

We also performed nonadiabatic computations for  white dwarf models
with  $M_{\star}= 0.666{\,}\rm M_{\sun}$ and $M_{\star}= 0.738{\,}\rm
M_{\sun}$ coming from progenitors with $Z= 0.0001$.  For the
$0.666{\,}\rm M_{\sun}$ sequence, we still found that the   dipole
$g_1$ mode is unstable mostly due to the $\varepsilon$ mechanism,
although for a narrow range of effective temperatures (see the upper
left panel of Fig.\, \ref{Fig:logtlogg}).  However, for the
$0.738{\,}\rm M_{\sun}$ sequence, we   found that mode no longer
excited by the $\varepsilon$ mechanism.  Again, this is an expected
result, since no appreciable nuclear burning occurs in these models
\citep[see][]{2015A&A...576A...9A}, and because the progenitor star of
this sequence experienced carbon enrichment in the AGB phase due to
third dredge-up episodes, as revealed in   the carbon to oxygen ratio
listed in Table{\,}\ref{table}.  We have estimated the expected
threshold mass ($\rm M_T$) above which the epsilon mechanism is not
operative anymore. Specifically, we artificially reduce the nuclear
energy release in the $0.666 {\,}\rm M_{\sun}$ sequence down to a
value below which the epsilon mechanism is not able to excite the
$g_1$ pulsation mode, and then we use the corresponding nuclear
luminosity at which this occurs to interpolate between our $0.666
{\,}\rm M_{\sun}$ and $0.738 {\,}\rm M_{\sun}$ sequences to find $\rm
M_T$. The stellar mass value turns out to be $\rm M_T \approx 0.71
{\,}\rm M_{\sun}$.

We now examine which  regions of the models contribute to the driving and 
damping of a given pulsation mode.  Figure{\,}\ref{Fig:wf} shows
the Lagrangian perturbation of the temperature ($\delta T/T$) for the
dipole $g_1$ mode (upper left-hand panel) and the $g_3$ mode (lower
left-hand panel), respectively, corresponding to a $Z= 0.0001$,
$M_{\star}= 0.535{\,}\rm M_{\sun}$ template model at $T_{\rm eff}= 11\,000$\,K.  
For the $g_1$ mode,  the eigenfunction $\delta T/T$ has a local maximum at
$\log(1-M_r/M_{\star})\sim -3.5$, where the hydrogen-burning shell  
is located.  The $\varepsilon$ mechanism acts as a
``filter'' that provides substantial driving to those $g$ modes
that have a maximum of $\delta T/T$ in the region of the nuclear
burning shell. For the dipole $g_3$ mode,  on other hand, 
$\delta T/T$ has negligible values at the burning shell region, 
hence the $\varepsilon$ mechanism does not provide efficient driving
for this mode.  

It is useful to introduce the ``running work integral'' $W(r)$,
which represents the work done on the overlying layer by the sphere
of radius $r$ \cite[see][]{1989nos..book.....U}. Its surface value,
$W := W(R)$, gives the increase of the total energy over one period of
oscillation. If $W > 0$, the pulsation is gaining energy in one cycle,
which means that the mode is unstable and grows. Otherwise, if $W < 0$, 
the mode is losing energy and damps.  The derivative of $W(r)$ 
gives us information about the regions of driving and damping of the
oscillations. Regions in the star where ${\rm d}W(r)/{\rm d}r > 0$
will contribute to driving, and zones in which 
${\rm d}W(r)/{\rm d}r < 0$ will provide damping.  

We show ${\rm d}W(r)/{\rm d}r$ for the dipole $g_1$ mode in the middle
upper panel of Fig.\,\ref{Fig:wf} with black dashed lines
(red dashed lines) for the case when the $\varepsilon$ mechanism is
allowed  (inhibited) to operate. The $\varepsilon$ mechanism  provides
significant driving at the hydrogen burning shell region,  but substantial
damping occurs there when the mechanism is inhibited. As a result, the running
work integral $W(r)$ increases  at the burning shell region when the
$\varepsilon$ mechanism  operates (solid black line), while it
substantially decreases when  the mechanism is switched off (solid red
line). As can be seen in the right-hand upper panel of this figure, 
when $\varepsilon$ mechanism operates, $W(r)$  becomes positive at the
surface of the star, thanks to the driving contribution due to the
 $\kappa-\gamma$ mechanism, and the $g_1$ mode becomes globally unstable.
In contrast, when the $\varepsilon$ mechanism  is inhibited, the
driving contribution by the   $\kappa-\gamma$ mechanism cannot prevent that
$W(r)$ becomes negative at the surface,  making the $g_1$ mode
stable in these circumstances.  

On the other hand, for the dipole $g_3$ mode  (lower panels of 
Fig.{\,}\ref{Fig:wf}), no significant driving is provided at the hydrogen  
burning shell region. However, since substantial driving occurs in the
hydrogen-ionization zone due to the  $\kappa-\gamma$ mechanism, $W(r)$ becomes 
positive at the surface. Thus, the $g_3$ mode is unstable, 
regardless of whether nuclear burning is taken into account 
or not in the stability computations.

\section{Summary and conclusions}
\label{conclusions}

\cite{2015A&A...576A...9A} predicted that low mass white dwarfs ---
evolved from low metallicity progenitors that did not experience
carbon enrichment  due to third dredge-up episodes during the AGB
phase--- are born with hydrogen envelopes thick enough to develop
intense hydrogen shell burning at low luminosities ($\log(L/\rm
L_\sun) \lesssim -3$).  Following these results, in this paper, we
have investigated whether   $g$ modes in DA white dwarfs could be
unstable due to the  $\varepsilon$ mechanism powered by the hydrogen
burning shell.   For that purpose, we computed nonadiabatic pulsation
modes of white dwarf models with $M_{\star} \sim
0.51$\,-\,$0.54{\,}\rm M_{\sun}$, evolved   from progenitorial stars
with four different metallicities: $Z= 0.0001$,  $Z= 0.0005$, $Z=
0.001$, and $Z= 0.018$.  The first three sequences  correspond to
low-metallicity stellar populations, while the latter  one is
representative of objects in the solar neighborhood.  We have found
that for the white dwarf sequences evolved   from low-metallicity
progenitors, both dipole and quadrupole $g_1$ modes are unstable
mostly due to the $\varepsilon$ mechanism acting at the  hydrogen
burning shell. We found that the ability of this mechanism to
destabilize those pulsation modes decreases with increasing
metallicity.  In the limit of white dwarfs with solar metallicity
progenitors ($Z= 0.018$), no pulsation is driven  by the
$\varepsilon$ mechanism, since nuclear burning does not play a  role
in these objects.  We have also explored the mass dependence of the
$\varepsilon$ mechanism efficiency,  by computing white dwarf
sequences with higher masses ($M_{\star}= 0.666{\,}\rm M_{\sun}$ and
$M_{\star}= 0.738{\,}\rm M_{\sun}$) evolved from $Z= 0.0001$
progenitors.  Our calculations showed that the $\varepsilon$ mechanism
destabilizes both the dipole and quadrupole $g_1$ modes associated
with the $0.666{\,}\rm M_{\sun}$  sequence, but does not excite these
modes for the sequence  with $M_{\star}= 0.738\,\rm M_{\sun}$. 

To summarize,  hydrogen shell burning in DA white dwarfs  at effective
temperatures typical of the instability strip of ZZ Ceti  stars 
triggers  the dipole ($\ell= 1$) and the quadrupole ($\ell= 2$)
$g_1$ modes with period $\Pi \sim 70 - 120$\,s,
provided that the mass of the white dwarf  is $M_{\star}
\lesssim 0.71{\,}\rm M_{\sun}$ and the metallicity of the  progenitor stars
is in the range $0.0001 \lesssim Z \lesssim 0.001$.  The $e$-folding
times for these  modes are much shorter than the  evolutionary timescale,
which means that they could have enough  time to  
reach observable amplitudes.  We hope that, in the future, these
modes can be detected in white dwarfs in low-metallicity
environments, such as globular clusters
and/or the galactic halo,  allowing us to place constraints on nuclear
burning in the envelopes of low-mass white dwarfs with  
low-metallicity. Moreover, detection of these pulsation
modes could eventually  verify that low-metallicity AGB stars  do
not experience carbon enrichment due to third dredge-up episodes.
Once again, asteroseismology of white dwarfs can prove to be a
powerful tool to sound the interior of these ancient stars, and also
to constrain  the history of their progenitors.

\begin{acknowledgements}
We acknowledge the comments and suggestions of our referee, S. O. Kepler, that improved the original version of this paper.
Part of this work was supported by AGENCIA  through the Programa de
Modernizaci\'on
Tecnol\'ogica BID  1728/OC-AR, by  the PIP 112-200801-00940  grant from
CONICET. This research has made use of NASA Astrophysics Data System. 
\end{acknowledgements}

\bibliographystyle{aa} 
\bibliography{epsilon}

\end{document}